\documentclass[preprint,aps,12pt,preprintnumbers,eqsecnum,nofootinbib]{revtex4}
\usepackage{graphicx}
\usepackage{subfigure}
\newcommand{\dof}{\mathrm{d.o.f.}}
\usepackage{color}
\usepackage{amssymb,amsmath}

\begin{document}
%
%
\title{\vspace*{0.5in} On the Cosmic-Ray Spectra of  \\ Three-Body Lepton-Flavor-Violating Dark Matter Decays
\vskip 0.1in}
\author{Christopher D. Carone}\email[]{cdcaro@wm.edu}
\author{Ari Cukierman}\email[]{ajcukierman@email.wm.edu}
\author{Reinard Primulando}\email[]{rprimulando@email.wm.edu}
\affiliation{High Energy Theory Group, Department of Physics,
College of William and Mary, Williamsburg, VA 23187-8795}
\date{August 2011}
\begin{abstract}
We consider possible leptonic three-body decays of spin-$1/2$, charge-asymmetric dark matter.  Assuming a general Dirac structure for the four-fermion contact interactions of interest, we 
study the cosmic-ray electron and positron spectra and show that good fits to the current data can be obtained for both charged-lepton-flavor-conserving and flavor-violating decay 
channels. We find that different choices for the Dirac structure of the underlying decay operator can be significantly compensated by different choices for the dark 
matter mass and lifetime. The decay modes we consider provide differing predictions for the cosmic-ray positron fraction at energies higher than those currently probed at the
PAMELA experiment; these predictions might be tested at cosmic-ray detectors like AMS-02.
\end{abstract}
\maketitle

\section{Introduction} \label{sec:intro}

Cosmic rays have been studied extensively at various earth-, balloon- and satellite-based experiments. Recently, the PAMELA satellite has observed an unexpected rise in the cosmic-ray positron fraction from approximately $7$ to
$100$~GeV~\cite{Adriani:2008zr}. This feature is not explained by the expected background from the secondary production of cosmic-rays positrons. Moreover, observations of the total flux of electrons and positrons by 
Fermi-LAT~\cite{Abdo:2009zk} and H.E.S.S.~\cite{Aharonian:2009ah}  also show an excess over the predicted background, up to an energy of $\sim 1$~TeV. The presence of nearby pulsars could provide an astrophysical explanation 
for these observations~\cite{Hooper:2008kg, Yuksel:2008rf}. Nevertheless, more exotic scenarios remain possible.  The annihilation of dark matter in the galactic halo to electrons and positrons provides one such possibility, though
generic annihilation cross sections must be enhanced by a large boost factor in order to describe the data~\cite{Cholis:2008hb, Cirelli:2008pk}.  Alternatively, the excess could be explained by a TeV-scale decaying dark matter
candidate. (For example, see Ref.~\cite{new1}; for a recent review, see Ref.~\cite{Fan:2010yq}.) In this scenario, fits to the cosmic-ray data indicate that the dark matter must decay primarily to leptons with a lifetime of ${\mathcal O}(10^{26})$~s.

While the thermal freeze-out of weakly-interacting, electroweak-scale dark matter can naturally lead to the desired relic density, this is not the only possible framework that can account for the present dark matter abundance. 
Recently proposed asymmetric dark matter models relate the baryon or lepton number densities to the dark matter number density, motivated by the fact that these quantities are not wildly dissimilar~\cite{adm1,Buckley:2010ui,Shelton:2010ta,Falkowski:2011xh}.  TeV-scale asymmetric dark matter models have been constructed, for example, in Refs.~\cite{Buckley:2010ui,Shelton:2010ta,Falkowski:2011xh}.  The asymmetry between dark matter particles and antiparticles can lead to differences in the primary cosmic-ray spectra of electrons and positrons, with potentially measurable consequences~\cite{new2,Chang:2011xn}.  Evidence for such charge asymmetric dark matter decays would disfavor the pulsar explanation of the $e^\pm$  excess~\cite{Chang:2011xn}.   In addition, charge asymmetric dark matter decays may allow one to discern whether dark matter decays are lepton-flavor-violating~\cite{Masina:2011hu}.  For example, the cosmic-ray spectra that one expects if dark matter decays symmetrically to $e^+\mu^-$ and $e^-\mu^+$ are indistinguishable from those obtained by assuming flavor-conserving decays to  $e^+ e^-$ and $\mu^+ \mu^-$ with equal branching fraction;  the same is not true if  the dark matter decays asymmetrically to $e^+ \mu^-$ alone, 100\% of the time.  

Refs.~\cite{Chang:2011xn} and \cite{Masina:2011hu} study the cosmic-ray $e^\pm$ spectra assuming a number of two-body charge-asymmetric dark matter decays, with the latter work focusing on
lepton-flavor-violating modes.  In this paper, we extend this body of work to charge-asymmetric three-body decays and, in particular, to modes that violate lepton flavor.  We assume a spin-$1/2$ 
dark matter candidate that decays via four-fermion contact interactions to two charged leptons and a light, stable neutral particle.  For the present purposes, the latter could either be a standard
model neutrino or a lighter dark matter component. Four-fermion interactions have a long history in the development of the weak interactions, and one can easily imagine that dark matter decays could be the consequence of operators of this form, generated by higher-scale physics.  Moreover, the possible presence of a neutrino in the primary decay may lead to interesting signals at neutrino telescopes~\cite{nuscopes}.  Unlike the two-body decays already considered in the literature, the precise energy distribution of the decay products is affected by the Dirac matrix structure of these contact interactions, which is not known (unless a model is  specified).    By considering the most general possibilities, we show that different choices for the Dirac structure of the decay operators defined in Sec.~\ref{sec:ops} can be substantially compensated by different choices for the dark matter mass $m_\psi$ and lifetime $\tau_\psi$; while the best fit values of these parameters change, the predicted spectra are not dramatically altered.  On the other hand, we find that the flavor structure of the decay operator has a more significant effect.   Assuming various lepton-flavor-conserving and flavor-violating decay modes, we compute the resulting cosmic-ray spectra, performing  $\chi^2$ fits to the data to determine the optimal dark matter masses and lifetimes.  Like Refs.~\cite{Chang:2011xn,Masina:2011hu}, we obtain predictions for these spectra at $e^\pm$ energies that are higher than those than can be probed accurately now.   Future data from experiments like AMS-02~\cite{amsx} may provide the opportunity to test these predictions, and evaluate them relative to other interpretations of the cosmic-ray positron excess.

This letter is organized as follows. In the next section, we discuss the assumed form of the dark matter operators. In Sec.~\ref{sec:spectra}, we present the results of our numerical analysis and in Sec.~\ref{sec:conc}, we discuss our results and directions for future work.

\section{Four-Fermion Operators} \label{sec:ops}

We consider a spin-$1/2$ dark matter candidate $\psi$ that decays to $\ell^+_i \ell^-_j \nu$ where $i$ and $j$ are generation indices and $\nu$ represents a light, neutral particle.  
We assume that $\nu$ is either a standard model neutrino or a secondary dark matter component that is much lighter than $\psi$ and contributes negligibly to the relic density.  In the 
present analysis, the exact nature of the light neutral state will be irrelevant since its effect on our results will come solely from kinematics.  We focus on the simplest scenario, in which 
there are no additional decay channels involving the charge conjugate of $\nu$, and  consider the possible four-fermion operators that contribute to the decays of interest.  We work directly with the operators that may appear after 
the standard model electroweak gauge symmetry is spontaneously broken; for any operator found to have phenomenologically desirable properties, one may easily construct a gauge-invariant origin after the fact.  Note that the
production of a neutrino in the primary decay may have interesting phenomenological consequences (see, for example, Ref.~\cite{nuscopes}), which provides a separate motivation for our three-fermion
final state.  Once this choice is made, the dark matter spin must be $1/2$ if the underlying theory is renormalizable~\footnote{For a model with flavor-conserving, three-body decays involving a 
final-state gravitino, see Ref.~\cite{ianL}.}.

The problem of parametrizing an unknown decay amplitude of one spin-$1/2$ particle to three distinct spin-$1/2$ decay products was encountered in the study of muon decay, before the
standard model was well established.  The most general decay amplitude ${\mathcal M}$ can be parametrized by~\cite{candb}
\begin{equation}
i {\mathcal M} = i g \sum_i \left[ \overline{u}(p_0)  O_i u_\psi \right] \,  \left[ \overline{u}(p_-) O_i (c_i + c_i^\prime \gamma^5) v(p_+) \right] \, ,
\label{eq:genamp}
\end{equation}
where $p_\pm$ and $p_0$ are the momenta of the decay products, labeled according to their electric charge, and the $O_i$, $i=1 \cdots 5$ are elements of the set of linearly independent
matrices
\begin{equation}
O = \{ 1 ,\, \gamma^\mu,\, \sigma^{\mu\nu},\, \gamma^\mu \gamma^5,\, \gamma^5\} \, .
\end{equation}
The $c_i$ and $c_i^\prime$ are complex coefficients.   Terms involving the contraction of spinor indices that link different pairs of spinor wave functions can be recast in the form of Eq.~(\ref{eq:genamp}) via Fierz transformations. Since the final state particles are much lighter than the dark matter candidate (which is at the TeV scale), we can safely neglect their masses.   

Since the neutral final state particle is stable, the energy spectra of electrons and positrons that are observed at cosmic-ray observatories are determined by the energy spectra of the the charged leptons, $\ell^+$ and $\ell^-$, that are produced in the primary decay; this follows from the differential decay distribution
\begin{equation}
\frac{1}{\Gamma} \frac{d^2 \Gamma}{dE_0 dE_\pm}  = \frac{1}{64 \pi^3 m_\psi}  
\langle |{\mathcal M}|^2 \rangle \, ,
\label{eq:ddifdist}
\end{equation}
where $\langle |{\mathcal M}|^2 \rangle$ is the spin-summed/averaged squared amplitude.  We evaluate this quantity exactly from Eq.~(\ref{eq:genamp}) using FeynCalc~\cite{feyncalc}, and compute the $\ell^\pm$ energy distribution by integrating over the neutral lepton energy $E_0$.   We find that the result contains terms quadratic and cubic in $E_\pm$;  however, since the distribution must be normalized to unity,  the result has the following simple parametrization:
\begin{equation}
\frac{1}{\Gamma} \frac{d \Gamma}{d E_\pm} = 
\frac{1}{m_\psi} \frac{E_\pm^2}{m_\psi^2} \left[ \xi_\pm +\left(64-\frac{8}{3} \xi_\pm\right) 
\frac{E_\pm}{m_\psi} \right] \, .
\label{eq:sdifdist}
\end{equation}
The requirement that this expression remains positive over the kinematically accessible range $0 \leq E_\pm \leq m_\psi/2$ restricts the parameters $\xi_+$ and $\xi_-$ to fall within the range
\begin{equation}
0 \leq \xi_\pm \leq 96 \, .
\end{equation}
The $\xi_\pm$ are generally complicated functions of the operator coefficients $c_i$ and $c_i^\prime$; we provide these in the appendix. In the present analysis, however, the exact relations are not particularly
important; by leaving $m_\psi$ and $\tau_\psi$ as fitting parameters, one obtains very similar predicted spectra, independent of the choice of the $\xi_\pm$. The fact that {\em some} solution exists for any desired Dirac structure of the underlying four-fermion operator makes it potentially easier to construct explicit models.  Though we reserve the task of model-building to future work, it is worth noting, for example, that the operator
\begin{equation}
O^{RR}_{ij} \equiv  \overline{\nu} \gamma^\mu (1+\gamma^5) \psi \overline{\ell}_i \gamma_{\mu} (1+\gamma^5) \ell_j \, ,
\end{equation}
corresponding to $\xi_+=96$ and $\xi_-=48$, is a particularly interesting choice, since it is already gauge invariant under the standard model gauge group and may provide a simple starting point for constructing a plausible ultraviolet completion.

We computed the electron and positron spectra using PYTHIA~\cite{pythiaman}, taking into account the energy distributions of the primary leptons $\ell^+$ and $\ell^-$.  As a cross check, we have written code that incorporates Eq.~(\ref{eq:ddifdist}), computed directly from a choice of the underlying four-fermion operator, as well as code that incorporates only the distributions 
Eq.~(\ref{eq:sdifdist}), for the corresponding values of $\xi_+$ and $\xi_-$.  We have also compared output from different versions of our code, based on PYTHIA 6.4 and PYTHIA 8.1,
respectively\footnote{Note that PYTHIA 6.4 does not automatically take into account neutron decay, which we include by modifying the program's decay table.}.  Results from these different approaches were found to be agreement.

\section{Cosmic-Ray Spectra}\label{sec:spectra}

To compute the relevant cosmic-ray fluxes, one must take into account that electrons and positrons produced in dark matter decays must propagate through the
galaxy before reaching earth.   While modeling this propagation is now standard in the literature on decaying dark matter scenarios, we briefly summarize our approach so that our
discussion is self contained and our assumptions are manifest.

\subsection{Cosmic-Ray Propagation}

Let $\mathbf{r}$ be a position with respect to the center of the Milky Way Galaxy.  We assume the spherically symmetric Navarro-Frenk-White dark matter halo density profile~\cite{navarro}
\begin{equation}
\rho ( r ) = \rho_0 \frac{r_c^3}{r (r + r_c)^2} \, ,
\end{equation}
where $\rho_o \simeq 0.26$ GeV/cm$^3$ and $r_c \simeq 20$ kpc.  The production rate of electrons/positrons per unit energy and per unit volume is then given by
\begin{equation}
Q(E, r) = \frac{\rho(r)}{m_\psi} \left ( \frac{1}{\tau_{\psi}} \frac{dN_{e^{\pm}}}{dE} \right ) \, ,
\end{equation}
where $m_\psi$ and $\tau_\psi$ are the dark matter mass and lifetime, respectively, and $dN_{e^{\pm}}/dE$ is the energy spectrum of electrons/positrons produced in the dark matter decay.  Let $f_{e^{\pm}}(E, \mathbf{r})$ be the number density of electrons/positrons per unit energy.  Then, $f_{e^{\pm}}(E, \mathbf{r})$ satisfies the transport equation~\cite{transport eqn}
\begin{equation}
 0 = K(E) \nabla^2 f_{e^{\pm}}(E, \mathbf{r}) + \frac{\partial}{\partial E} \left [ b(E) f_{e^{\pm}}(E, \mathbf{r}) \right ] + Q(E,r).
\end{equation}
We assume the MED propagation model described in Ref.~\cite{MED} for which
\begin{equation}
K(E) = 0.0112 \epsilon^{0.70} \, \mathrm{kpc}^2 / \mathrm{Myr}
\end{equation}
and
\begin{equation}
b(E) = 10^{-16} \epsilon^2 \, \mathrm{GeV}/\mathrm{s} \, ,
\end{equation}
where $\epsilon = E / (1 \, \mathrm{GeV})$.  The diffusion zone is approximated as a cylinder with half-height $L = 4$ kpc and radius $R = 20$ kpc.  We require $f_{e^{\pm}}(E, \mathbf{r})$ to vanish at the boundary of this zone.  The solution at the heliospheric boundary is then given by~\cite{Green's function}
\begin{equation}
f_{e^{\pm}}(E) = \frac{1}{m_\psi \tau_{\psi}} \int\limits_0^{m_\psi} dE' \, G_{e^{\pm}} (E, E') \frac{dN_{e^{\pm}} (E')}{dE'}.
\end{equation}
The Green's function, $G_{e^{\pm}}(E,E')$, can be found in Ref.~\cite{Green's function}.
The interstellar flux of electrons/positrons created in dark matter decays is then given by
\begin{equation}
\Phi^{\mathrm{DM}}_{e^{\pm}}(E) = \frac{c}{4 \pi} f_{e^{\pm}}(E) \, ,
\end{equation}
where $c$ is the speed of light.

For the background fluxes, we assume the Model 0 proposed by the Fermi collaboration~\cite{Model 0,solmod}:
\begin{equation}
\Phi^{\mathrm{bkg}}_{e^-}(E) = \left ( \frac{82.0 \epsilon^{-0.28}}{1 + 0.224\epsilon^{2.93}} \right ) \, \mathrm{GeV}^{-1} \mathrm{m}^{-2} \mathrm{s}^{-1} \mathrm{sr}^{-1}
\end{equation}
and
\begin{equation}
\Phi^{\mathrm{bkg}}_{e^+}(E) = \left ( \frac{38.4 \epsilon^{-4.78}}{1 + 0.0002 \epsilon^{5.63}} + 24.0 \epsilon^{-3.41} \right ) \, \mathrm{GeV}^{-1} \mathrm{m}^{-2} \mathrm{s}^{-1} \mathrm{sr}^{-1} \, ,
\end{equation}
where, as before, $\epsilon = E / (1 \, \mathrm{GeV})$.

At the top of the Earth's atmosphere, these fluxes must be corrected to account for the effects of solar modulation~\cite{solmod}.  The flux at the top of the atmosphere (TOA) is related to the interstellar (IS) flux by
\begin{equation}
\Phi^{\mathrm{TOA}}_{e^{\pm}} (E_{\mathrm{TOA}}) = \frac{E_{\mathrm{TOA}}^2}{E_{\mathrm{IS}}^2} \Phi^{\mathrm{IS}}_{e^{\pm}} (E_{\mathrm{IS}}) \, ,
\end{equation}
where $E_{\mathrm{IS}} = E_{\mathrm{TOA}} + |e|\phi_F$ and $|e|\phi_F = 550 \, \mathrm{MeV}$.

The total electron-positron flux is given by 
\begin{equation}
\Phi^{\mathrm{tot}}_e = \Phi^{\mathrm{DM}}_{e^-}(E) + \Phi^{\mathrm{DM}}_{e^+}(E) + k \Phi^{\mathrm{bkg}}_{e^-}(E) + \Phi^{\mathrm{bkg}}_{e^+}(E) \, ,
\end{equation}
where $k$ is a free parameter which determines the normalization of the background electron flux.   In our numerical analysis, we find that the best fit values
of $k$ never deviate by more that two percent from $0.84$  and that fixing $k$ at this value has a negligible effect on the goodness of fits and our predicted spectra.
Therefore, we  set  $k = 0.84$  henceforth to reproduce the cosmic-ray spectra at low energies.  The positron fraction is given by
\begin{equation}
\mathrm{PF}(E) = \frac{\Phi^{\mathrm{DM}}_{e^+}(E)  + \Phi^{\mathrm{bkg}}_{e^+}(E)}{\Phi^{\mathrm{tot}}_e}.
\end{equation}

\subsection{Results}

In the propagation model described above, the only remaining undetermined quantities are $m_\psi$, $\tau_{\psi}$, $d N_{e^+} / d E$ and $d N_{e^-} / d E$.  The electron and positron energy spectra, $d N_{e^+} / d E$ and $d N_{e^-} / d E$, are determined by $m_\psi$ and by a set of parameters which we describe in the following paragraph.  

We consider dark matter decays of the form $\psi \rightarrow \ell_i^+ \ell_j^- \nu$ where $\ell_i^{\pm}$ is a charged lepton of the $i^{{\rm th}}$ generation.  There are nine such decay channels, and we require
\begin{equation}
\label{Bsum}
\displaystyle \sum_{i,j} B(\ell_i^+ \ell_j^- \nu) = 1 \, ,
\end{equation}
where the $B(\ell_i^+ \ell_j^- \nu)$ are branching fractions.  For decays involving more than one channel,
\begin{equation}
\frac{d N_{e^{\pm}}}{d E} = \displaystyle \sum_{i,j} B(\ell^+_i \ell^-_j \nu) \left ( \frac{d N_{e^{\pm}}}{d E} \right )_{ij} \, ,
\end{equation}
where $\left ( d N_{e^{\pm}} / d E \right )_{ij}$ is the electron/positron energy spectrum for $\psi \rightarrow \ell_i^+ \ell_j^- \nu$.  In Sec.~\ref{sec:ops}, we showed that the energy spectra of the charged leptons in the decay $\psi \rightarrow \ell_i^+ \ell_j^- \nu$ are characterized by the ordered pair $(\xi_+, \xi_-)$, where $0 \leq \xi_{\pm} \leq 96$.  We also showed that $\left (d N_{e^{\pm}} / dE \right )_{ij}$ is entirely determined by $m_\psi$ and $(\xi_+, \xi_-)$.  For decays involving more than one decay channel (e.g., $\psi \rightarrow e^+ \mu^-\nu$ and $\psi \rightarrow \mu^+ \tau^-\nu$), we assume a constant $(\xi_+, \xi_-)$.  Then, since the branching fractions are subject to Eq.~(\ref{Bsum}), we can determine $d N_{e^+} / d E$ and $d N_{e^-} / d E$ by specifying $m_\psi$, $\xi_+$, $\xi_-$ and eight of the nine branching fractions.  

To summarize, when we use the cosmic-ray propagation model described in the previous subsection, the resulting positron fraction and total electron-positron flux measured at the top of the Earth's atmosphere are determined by $12$ parameters: $m_\psi$, $\tau_{\psi}$, $\xi_+$, $\xi_-$ and eight of the nine branching fractions.

For each of the decay scenarios considered below, we fixed $(\xi_+, \xi_-)$ and the branching fractions and then performed a $\chi^2$ fit to the PAMELA, Fermi LAT, H.E.S.S.~2008 and H.E.S.S.~2009 data with $m_\psi$ and $\tau_{\psi}$ as fitting parameters.  We allowed $m_\psi$ to vary in increments of $500$ GeV, and we allowed $\tau_{\psi}$ to vary in increments of $0.1 \times 10^{26}$ s.  We consider the range $E > 10$~GeV, where the effects of a TeV-scale dark matter candidate are relevant.  Where the high-energy and low-energy Fermi data overlap, we have plotted only the high-energy data. (We omit from our figures the H.E.S.S. bands of
systematic uncertainty.)

\begin{figure}
\subfigure{\includegraphics[width = 0.495\textwidth]{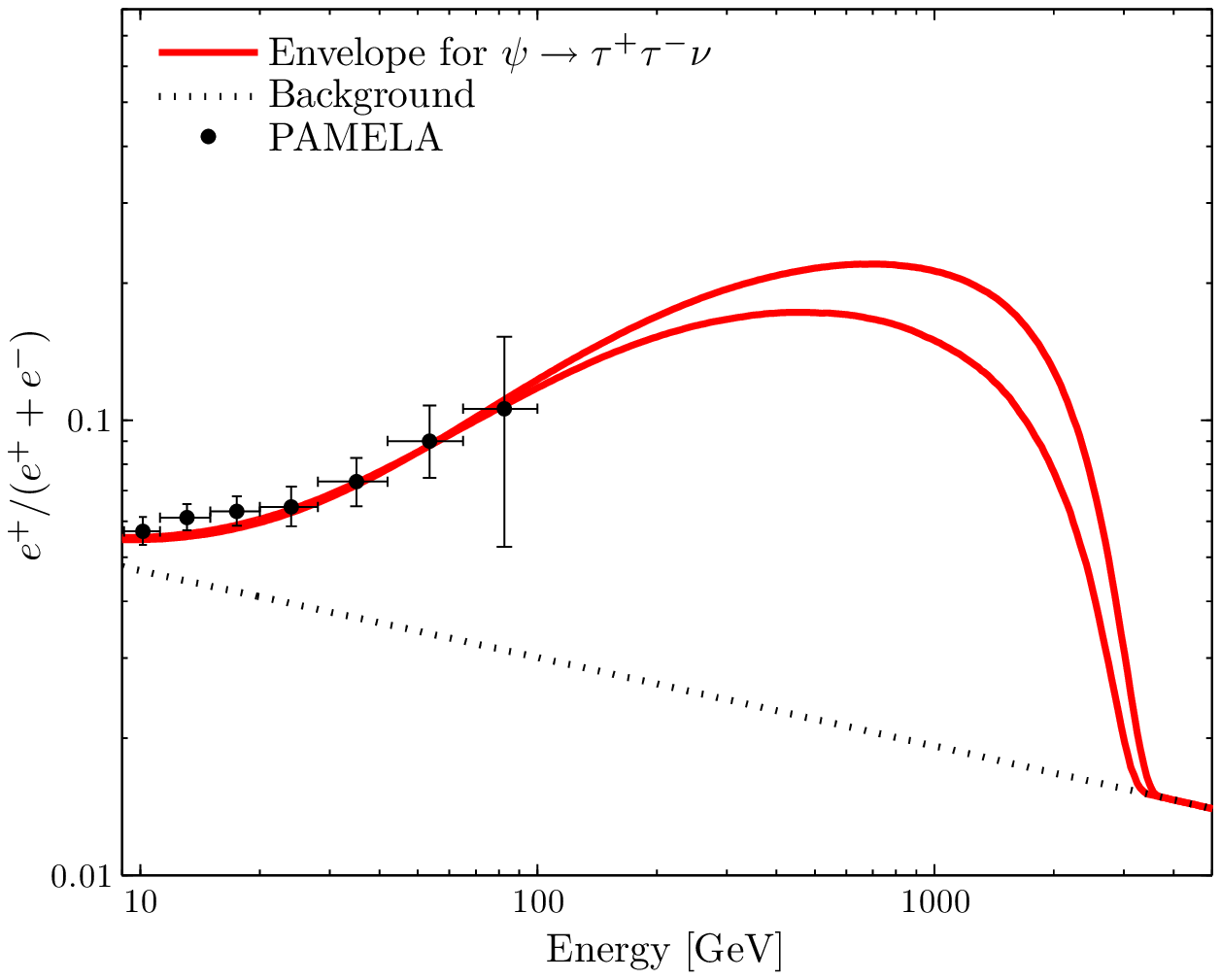}}
\subfigure{\includegraphics[width = 0.495\textwidth]{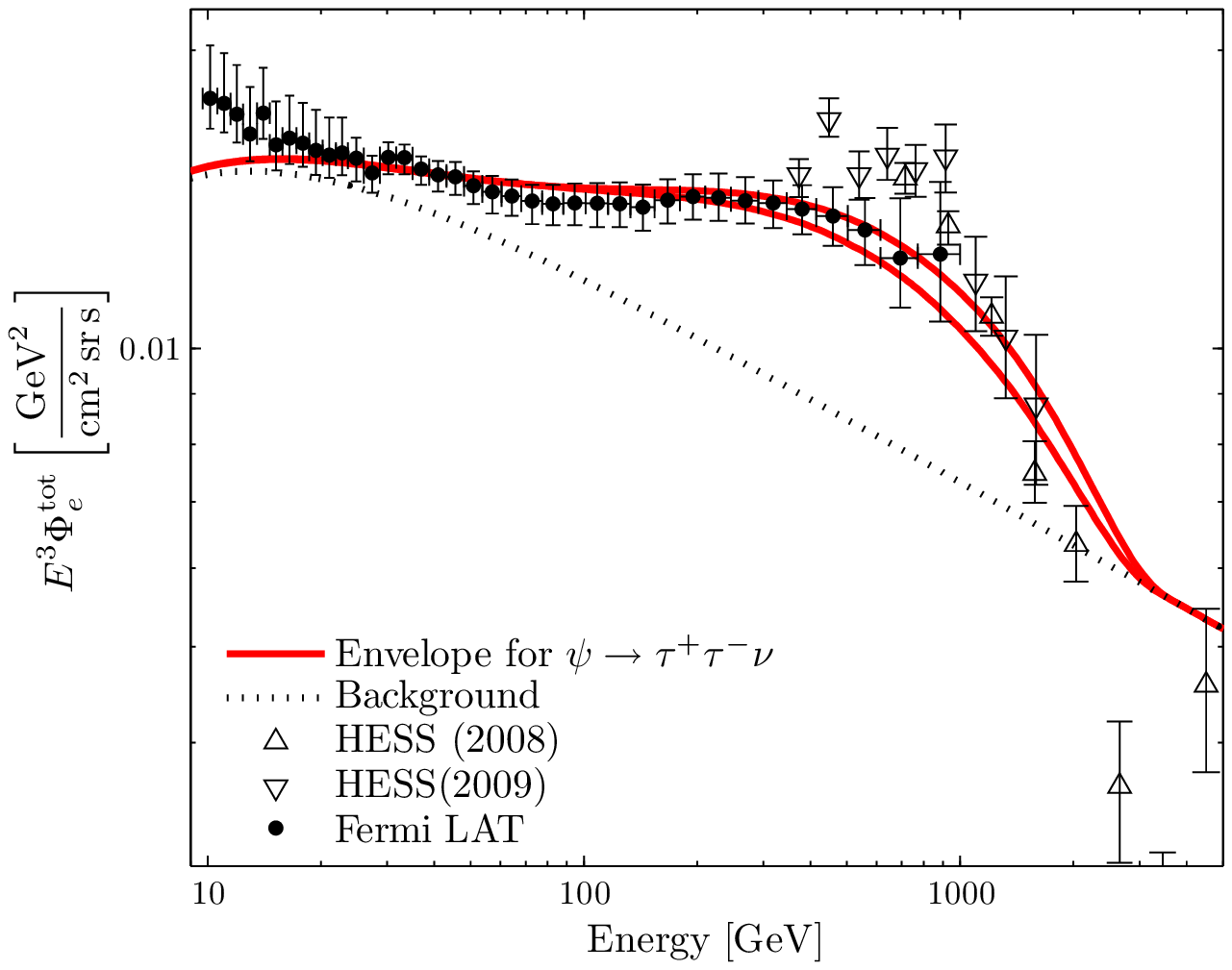}}
\caption{\label{tautau}The envelope of possible cosmic-ray spectra for $\psi \rightarrow \tau^+ \tau^- \nu$.  Ranges of the fit parameters are given in the text.}
\end{figure}

Leaving $m_\psi$ and $\tau_{\psi}$ as free variables, we find that our results are relatively insensitive to the choice of $(\xi_+, \xi_-)$.  This is demonstrated for the pure decay $\psi \rightarrow \tau^+ \tau^- \nu$ in Fig.~\ref{tautau} where we show the envelope of possible cosmic-ray spectra; that is, when we sample the $(\xi_+, \xi_-)$ parameter space, we find that all of the resulting curves fall between those plotted in Fig.~\ref{tautau}.  For the example shown,  $m_\psi$ varies between $6.5$ and $8.5$~TeV while
$\tau_\psi$ varies between $0.5 \times 10^{26}$~s and $0.7 \times 10^{26}$~s; the $\chi^2$ per degree of freedom ($\chi^2/\dof$) remains between $0.5$ and $0.6$.
We performed the same analysis on the other decay scenarios discussed below and found a similar behavior.  As such, we take $(\xi_+, \xi_-) = (48,48)$ for the remaining results that we 
present.

\begin{figure}
\subfigure{\includegraphics[width = 0.495\textwidth]{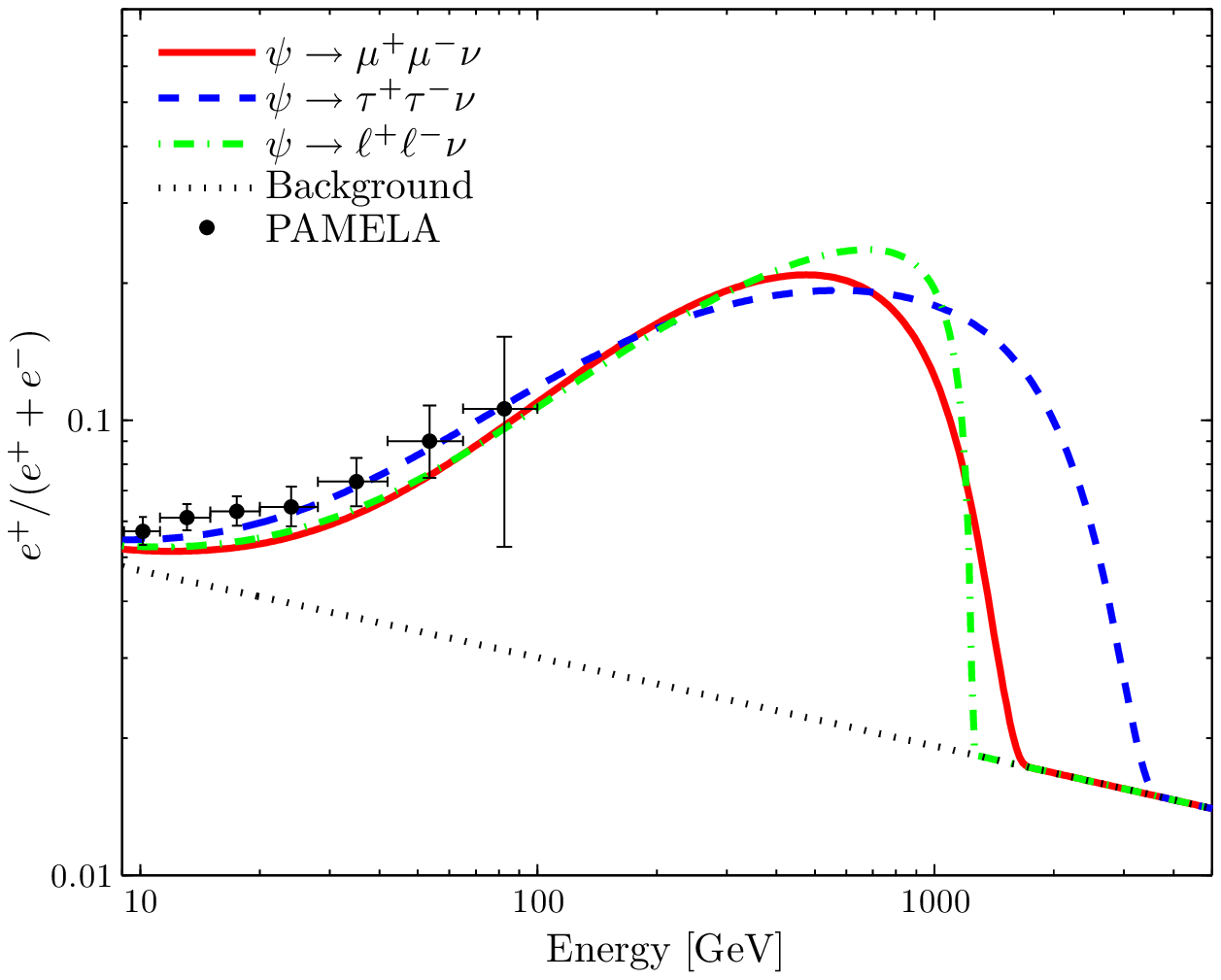}}
\subfigure{\includegraphics[width = 0.495\textwidth]{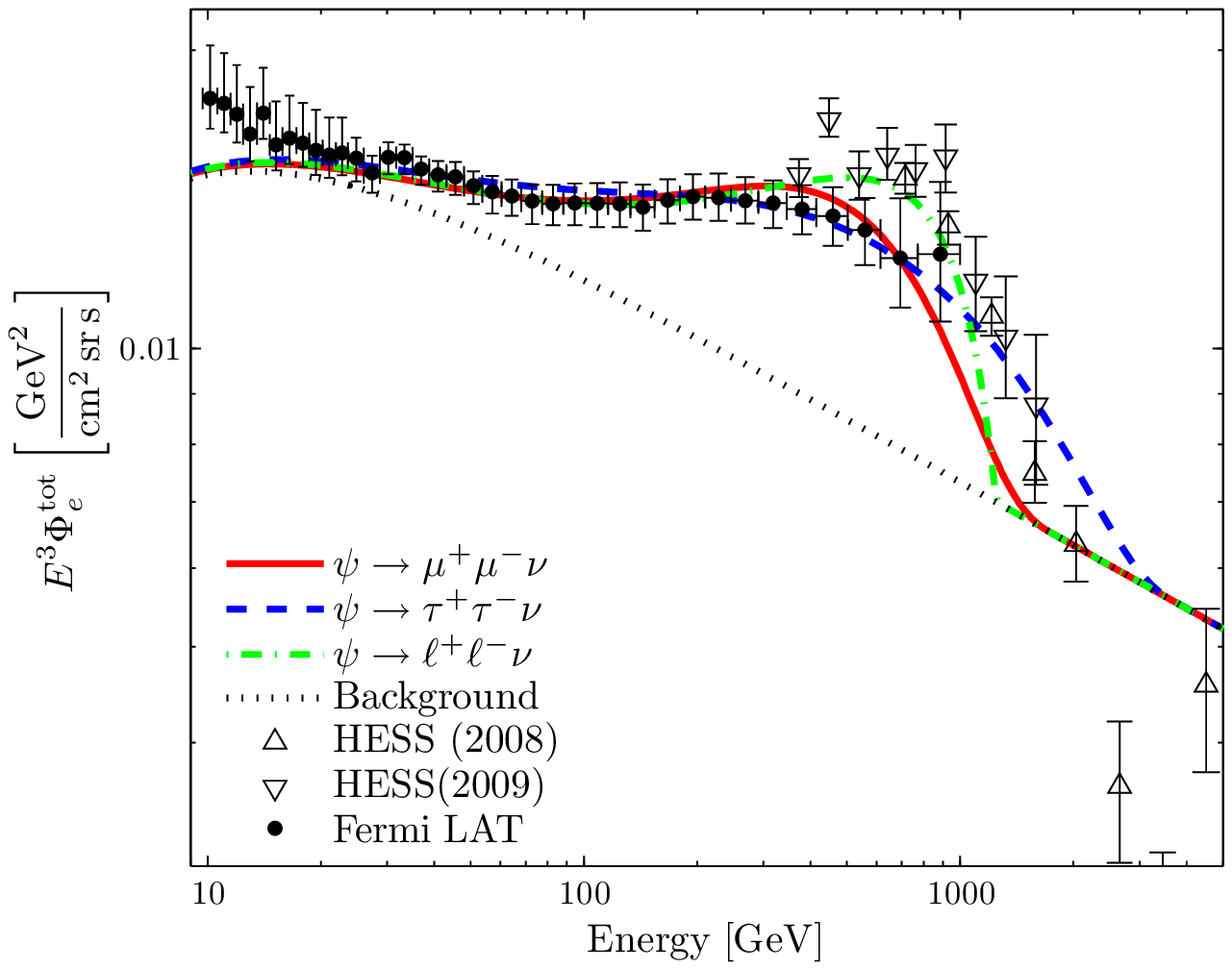}}
\caption{\label{FC} Positron fraction and total electron-positron flux for some charged-lepton-flavor-conserving decays.  Best fits are shown, corresponding to the following masses and lifetimes:
for $\psi \rightarrow \mu^+ \mu^- \nu$,  $m_\psi = 3.5$ TeV and $\tau_{\psi} = 1.5 \times 10^{26}$ s; for $\psi \rightarrow \tau^+ \tau^- \nu$,  $m_\psi = 7.5$ TeV and $\tau_{\psi} = 0.6 \times 10^{26}$ s;  for the 
flavor-democratic decay $\psi \rightarrow \ell^+ \ell^- \nu$,  $m_\psi = 2.5$ TeV and $\tau_{\psi} = 1.9 \times 10^{26}$ s.}
\end{figure}

\begin{figure}
\subfigure{\includegraphics[width = 0.495\textwidth]{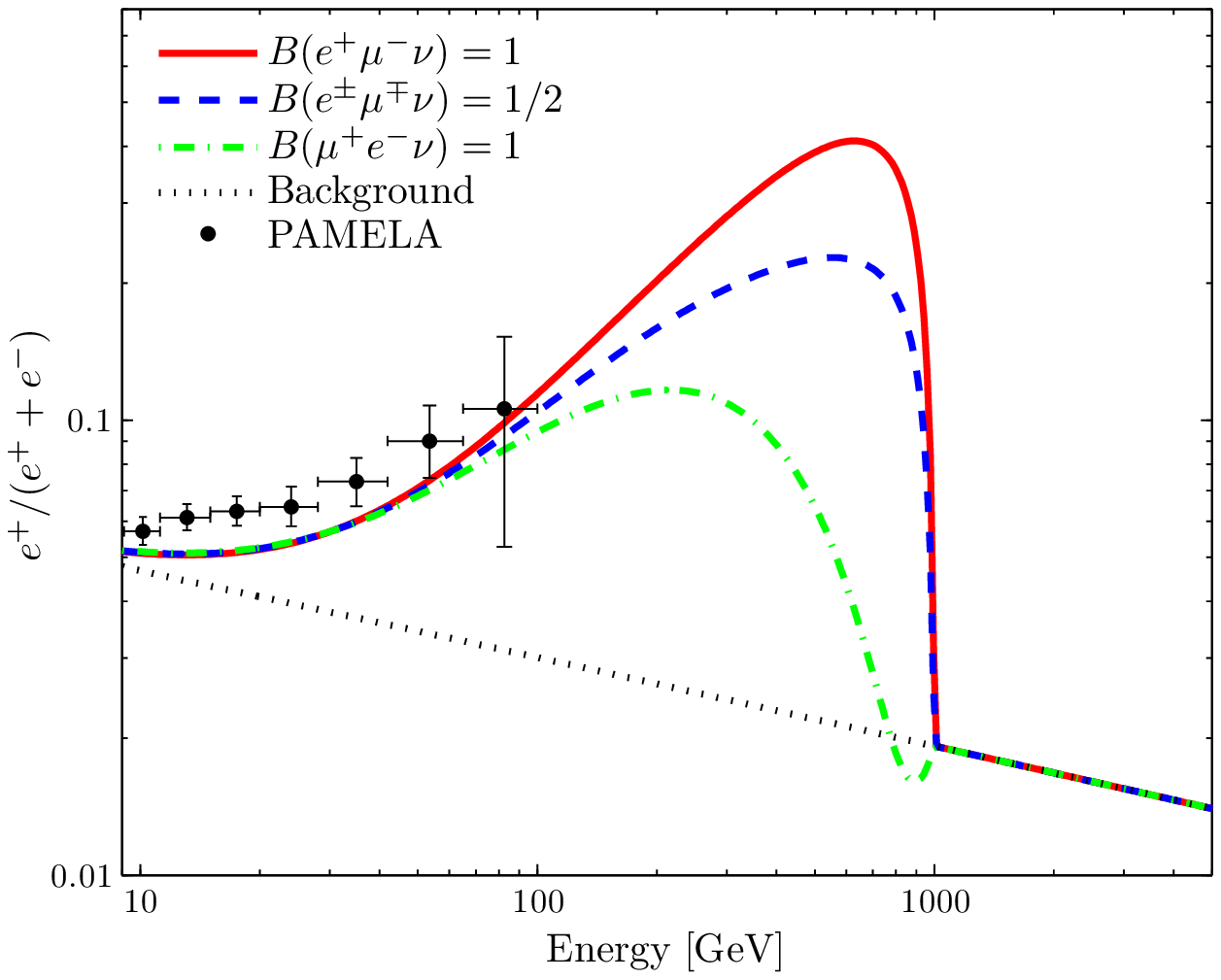}}
\subfigure{\includegraphics[width = 0.495\textwidth]{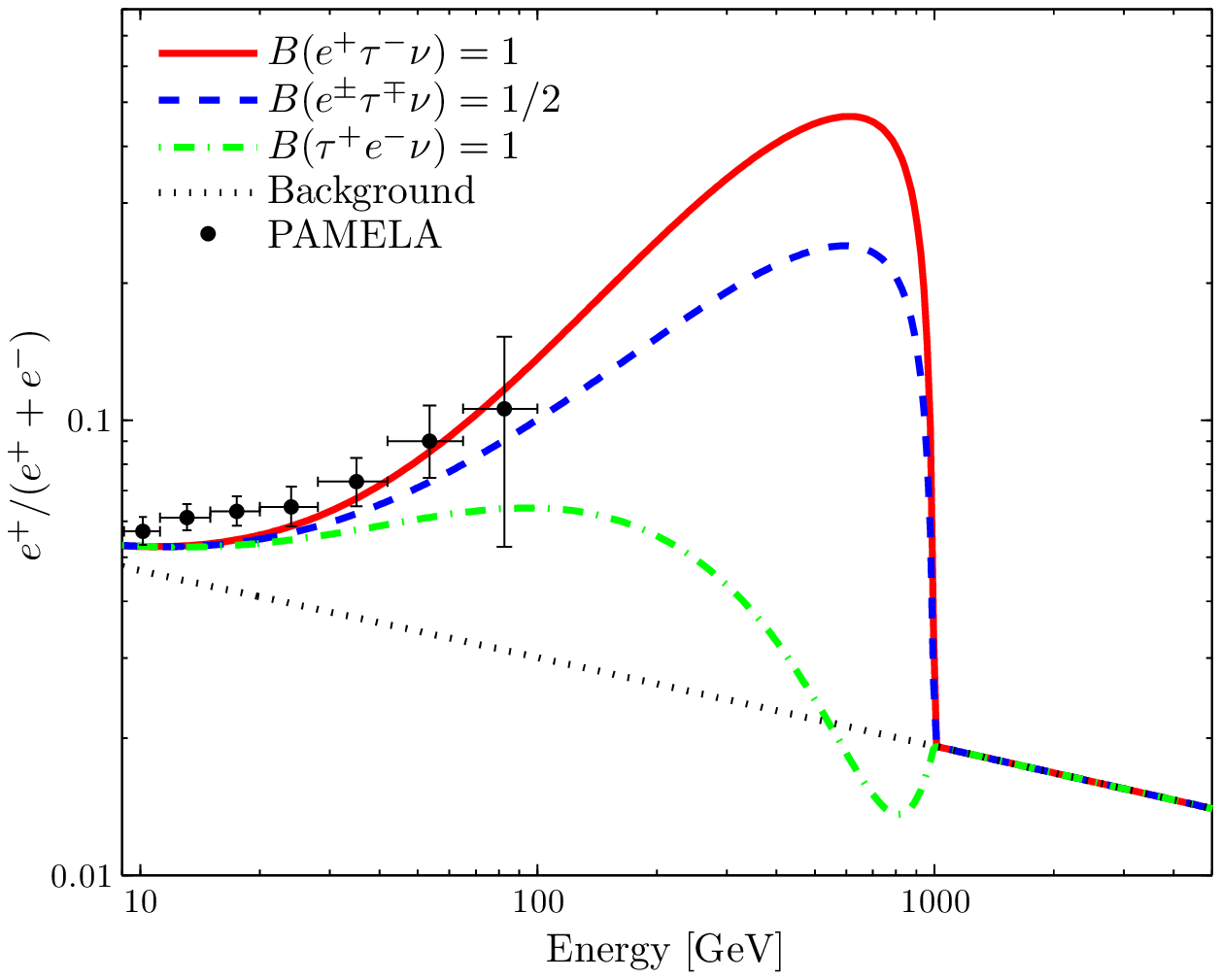}}
\subfigure{\includegraphics[width = 0.495\textwidth]{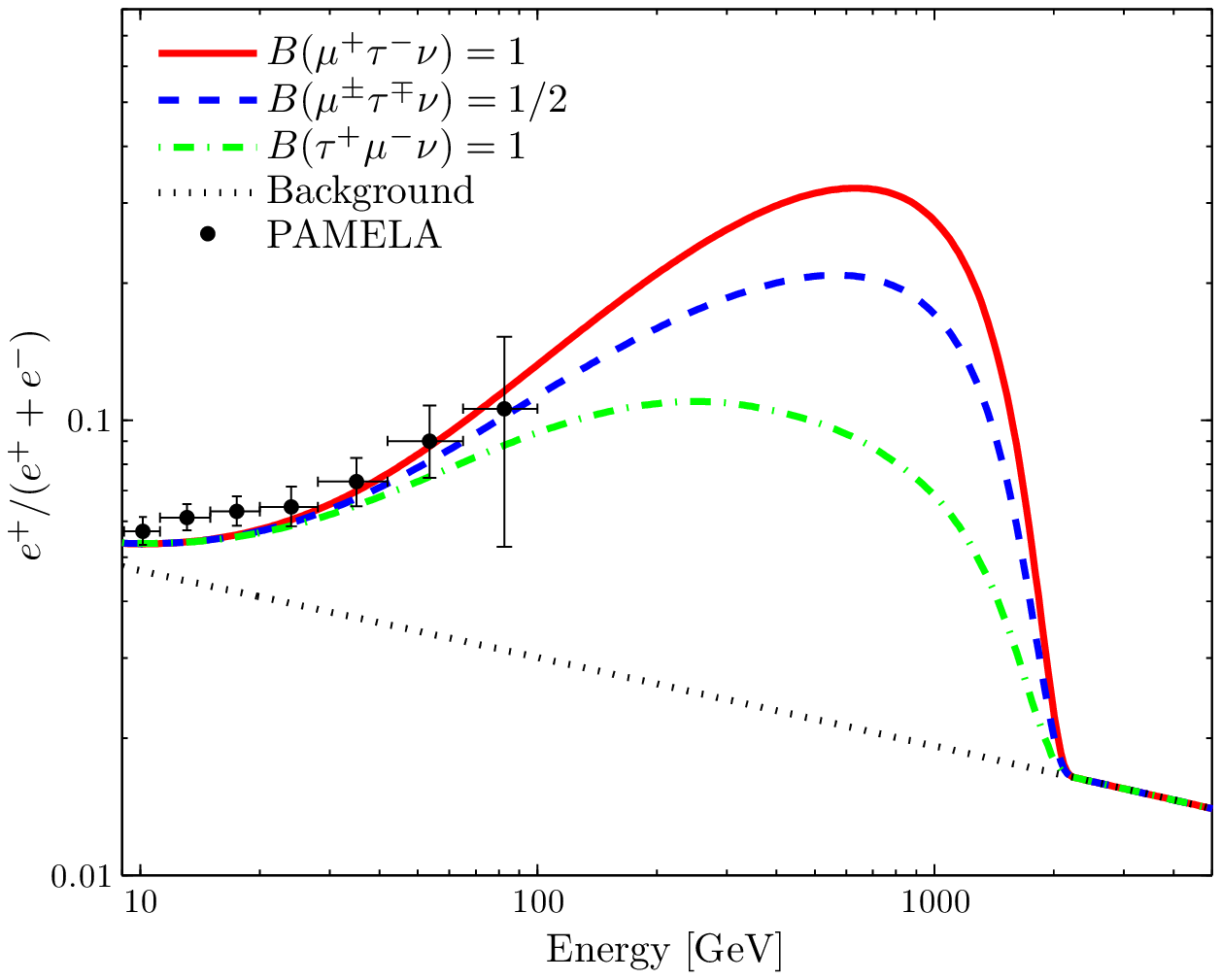}}
\subfigure{\includegraphics[width = 0.495\textwidth]{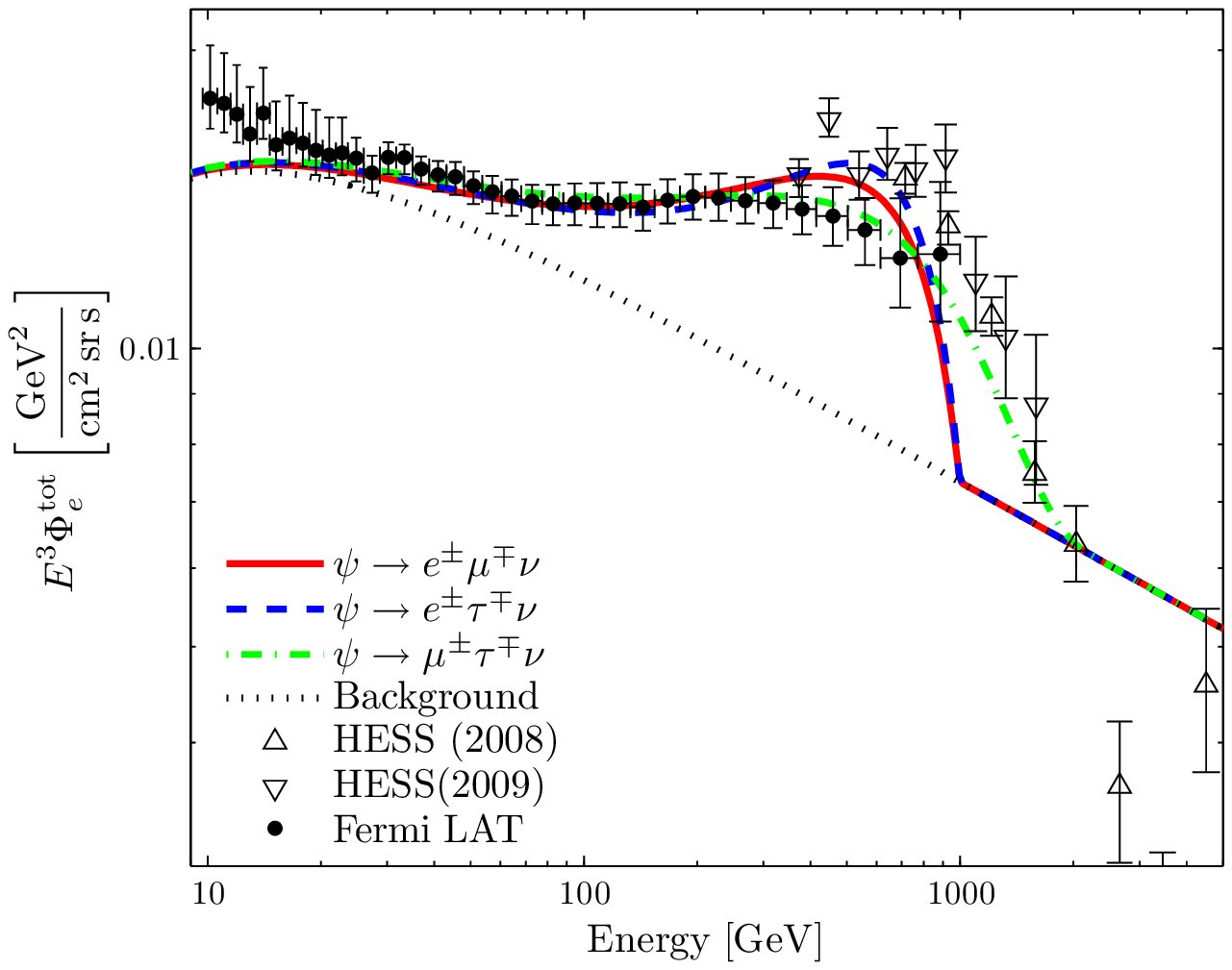}}
\caption{\label{FV}Positron fraction and total electron-positron flux for some charged-lepton-flavor-violating decays with various sets of branching fractions. 
Best fits are shown, corresponding to the following masses and lifetimes: for $\psi \rightarrow e^{\pm} \mu^{\mp} \nu$, $m_\psi = 2.0$ TeV and $\tau_{\psi} = 2.9 \times 10^{26}$ s;  
for $\psi \rightarrow e^{\pm} \tau^{\mp} \nu$, $m_\psi = 2.0$ TeV and $\tau_{\psi} = 2.4 \times 10^{26}$ s; for $\psi \rightarrow \mu^{\pm} \tau^{\mp} \nu$,  $m_\psi = 4.5$ TeV and $\tau_{\psi} = 1.0 \times 10^{26}$ s.}
\end{figure}

As a starting point,  we show the cosmic-ray spectra for some charged-lepton-flavor-conserving decays in Fig.~\ref{FC}.
We consider the pure decays $\psi \rightarrow \mu^+ \mu^- \nu$ and $\psi \rightarrow \tau^+ \tau^- \nu$, and we also consider the flavor-democratic decay for which $B(\ell_i^+ \ell_i^- \nu) = 1/3$ for all $i$.  For $\psi \rightarrow \mu^+ \mu^- \nu$, we have a 
$\chi^2/\dof$ of approximately $0.9$.  For $\psi \rightarrow \tau^+ \tau^- \nu$, we have $\chi^2/\dof  \approx 0.6$.  And for the flavor-democratic $\psi \rightarrow \ell^+ \ell^- \nu$, we have $\chi^2/\dof \approx 0.8$.  These are to be contrasted with the flavor-violating decays of Fig.~\ref{FV}.  

We consider three classes of flavor-violating decays:
\begin{align}
\psi & \rightarrow e^{\pm} \mu^{\mp} \nu, & \psi & \rightarrow e^{\pm} \tau^{\mp} \nu, & & \mathrm{and} & \psi & \rightarrow \mu^{\pm} \tau^{\mp} \nu.
\end{align}
Each class contains two decay channels (e.g., $\psi \rightarrow e^+ \mu^- \nu$ and $\psi \rightarrow e^- \mu^+ \nu$).  We consider all six of the pure decays, i.e., decays involving only one channel.  We also consider mixtures of decay channels belonging to the same class; some representative choices are shown in Fig.~\ref{FV}.  
Note that, for fixed $m_\psi$ and $\tau_{\psi}$, the total electron-positron flux -- which does not distinguish between the two electric charges -- is the same for any two decays belonging to the same class.  For this reason, we require only one plot of the total flux in Fig.~\ref{FV}.  We find that  the $\chi^2$ is relatively flat as a function of the branching fraction within each class of decays:  over the range of possible branching fractions, we find that the $\chi^2/\dof$ varies by no more than $\pm 10\%$ from $1.2$, $1.1$ and $0.6$, for $\psi \rightarrow e^{\pm} \mu^{\mp} \nu$, $\psi \rightarrow e^{\pm} \tau^{\mp} \nu$, and  $\psi \rightarrow \mu^{\pm} \tau^{\mp} \nu$, respectively.  Different choices for the branching fraction within a given class describe the existing data well, but provide different predicted spectra that interpolate between the curves shown. Note that the distinctive dip in the $\mu^+ e^- \nu$ and $\tau^+ e^- \nu$ positron fractions around $1$~TeV is due to the hard electron produced in the initial decay; this greatly enhances the electron to positron ratio in the high energy bins, leading to a suppression in the positron fraction for fixed total flux.

\section{Discussion}\label{sec:conc}

The results presented in the previous section show that a variety of possible lepton-flavor-violating decay modes for a spin-$1/2$, charge asymmetric dark matter candidate can describe existing data well, as quantified 
by the $\chi^2$ per degree of freedom for the best fits to the data.   Significantly, the results for the predicted positron fraction differ substantially for energies above $\sim 100$~GeV, the maximum for which the PAMELA 
experiment is sensitive.  In some case,  more precise measurement of the total electron-postron flux around $1$~TeV may also provide a means of distinguishing these scenarios.   Future data from experiments like 
AMS-02~\cite{amsx}, which can probe these energy ranges of the predicted spectra, may determine whether the possibilities discussed in this letter present viable descriptions of the cosmic-ray spectrum.

In the meantime, the present work suggests a number of directions for further study:  In the case where the stable, neutral particle in the final state is a standard model neutrino, one could study whether the decays of asymmetric dark matter that we have considered could be probed at neutrino observatories like IceCube~\cite{nuscopes} .  One could also study additional astrophysical bounds on the scenarios described, for example, from the
extragalactic gamma ray flux~\cite{Chang:2011xn}.  One can also attempt to find preferred forms of the underlying four-fermion operators (whose effects were parametrized in the present analysis by $\xi_\pm$) by studying the simplest and best-motivated models that provide for their origin.  Work in these directions is in progress and will be described in a longer publication.

\begin{acknowledgments}
This work was supported by the NSF under Grants PHY-0757481 and PHY-1068008.  A.C.'s undergraduate research was supported, in part, by a William \& Mary Honors Fellowship.
C.D.C. gratefully acknowledges support from a William \& Mary Plumeri Fellowship.
\end{acknowledgments}

\appendix
\section{The Parameters $\xi_\pm$}

The parameters $\xi_\pm$ may be  expressed in terms of the operator coefficients $c_i$ and $c_i^\prime$ defined in Eq.~(\ref{eq:genamp}),
\begin{equation}
\xi_\pm = 48 \,\frac{ {\bf c}^\dagger N_\pm\, {\bf c} + {\bf c^\prime}^\dagger N_\pm \,{\bf c^\prime}}{{\bf c}^\dagger D \,{\bf c} + {\bf c^\prime}^\dagger D\, {\bf c^\prime}}  \, ,
\end{equation}
where ${\bf c}=[c_1,c_2,c_3,c_4,c_5]^T$ and ${\bf c^\prime} = [c_1^\prime,c_2^\prime,c_3^\prime,c_4^\prime,c_5^\prime]^T$.  The five-by-five matrices $N_\pm$ and $D$ are given by
\begin{equation}
N_\pm = \left(\begin{array}{ccccc} 
1 & 0 & \mp 2 & 0 & 0 \\
0 & 6 & 0 & \pm 2 & 0 \\
\mp 2 & 0 & 40 & 0 & \mp 2 \\
0 & \pm 2 & 0 & 6 & 0 \\
0 & 0 & \mp 2 & 0 & 1\end{array}\right)
\,\,\,\,\, \mbox{ and } \,\,\,\,\,
D = \left(\begin{array}{ccccc}
1 & 0 & 0 & 0 & 0 \\
0 & 4 & 0 & 0 & 0 \\
0 & 0 & 24 & 0 & 0 \\ 
0 & 0 & 0 & 4 & 0 \\
0 & 0 & 0 & 0 & 1 \end{array}\right) \,.
\end{equation}

\end{document}